# A Real-Time Voice Activity Detection Based On Lightweight Neural Network

*Jidong Jia[1*], Pei Zhao[1], Di Wang[1,2]*

1. Haier Smart Home Co., Ltd.

2. State Key Laboratory of Digital Household Appliances, Qingdao

`jia_jidong@163.com`

## Abstract

Voice activity detection (VAD) is the task of detecting speech in an audio stream, which is challenging due to numerous unseen noises and low signal-to-noise ratios in real environments. Recently, neural network-based VADs have alleviated the degradation of performance to some extent. However, the majority of existing studies have employed excessively large models and incorporated future context, while neglecting to evaluate the operational efficiency and latency of the models. In this paper, we propose a lightweight and real-time neural network called MagicNet, which utilizes casual and depth separable 1-D convolutions and GRU. Without relying on future features as input, our proposed model is compared with two state-of-the-art algorithms on synthesized in-domain and out-domain test datasets. The evaluation results demonstrate that MagicNet can achieve improved performance and robustness with fewer parameter costs.

## 1. Introduction

It is a basic but important task to distinguish speech from noise in audio, named Voice activity detection (VAD), in many applications, such as speech and noise power spectrum estimation as well as automatic speech and speaker recognition. With the widespread use of digital home assistants, VAD is becoming challenging due to the wide variation of noise, low signal to noise ratio (SNR) and limited computing power in the scene of far-field voice interaction where the commands are spoken at a distance from the sound-capturing device[1].

Early research on VAD were some heuristics methods based on features of short-time energy, zero-crossing rate, duration and pitch as well as some methods of likelihood ratio test based on simple statistical model[2, 3]. Those approaches were limited in the condition of high SNR and stationary noise because of their simplistic features and impractical assumptions.

The rise of machine learning provided a new route for addressing the limitation that the performance of VAD degrade in low SNR environments[4, 5]. A typical method of machine learning was the joint of hidden Markov model and Gaussian mixture model (HMM-GMM), where two GMMs were trained on two classes of data respectively as a binary classifier to describe probability distributions of speech as well as non-speech and a HMM was viewed as a post filter to smooth probabilities estimated by the classifier for the temporal continuity of audio [6, 7].

During the last decade, a kind of more powerful classifier based on artificial neural network has been researched widely[8]. An impressive improvement in performance of VAD based on deep neural networks (DNN) has been made than one based on GMM due to the super modeling ability of DNN and extended context features [9]. Recurrent neural networks (RNN) specially designed for modeling sequences of inputs has been used in VAD task to optimize frame classification and probability dynamics simultaneously [10]. Although standard RNN are able to model a limited amount of temporal dependency, long-short term memory recurrent neural networks (LSTM) go beyond the limitation by introducing the concept of a memory cell and were used in VAD outperforming than state-of-the-art statistical VAD algorithms [11]. In [12], author think that LSTM suffer from state saturation problems when the utterance is long so proposed an architecture of convolution neural network (CNN) to model temporal variations. To study robustness of deep learning approaches, DNN, LSTM, CNN were compared and results demonstrated that LSTM was more robust than CNN and DNN under various circumstances[8].

Recently, a novel framework that consists of CNN, RNN and DNN has attracted more attentions because the combination of those networks could combines advantages of different neural networks [13-16]. However, most of them ignored that more future frames would cause more latency between input and output as well as storage and calculation resources of smart terminal devices are limited. Moreover, The performance of VAD was affected significantly by the vision on future data[10]. In the paper, we propose a light weight and low latency neural network used in VAD. It is inspired by the MobileNet and gated recurrent unit neural network (GRU) so we name it MagicNet. Here, we keep the real-time constraint on VAD means that only history samples could be used by networks to make the decision of whether the current audio frame is voice or noise. Meanwhile, the CNN layers is designed employing 1-D depth separable convolutions and residual structure to reduce the number of parameters in the model and makes it be a light neural network.

The rest of this paper is as follows. Section 2 introduces the proposed model. The detail and results of experiments are presented and discussed in Section 3. Conclusions are given in Section4.

## 2. Proposed Model Architecture

The proposed VAD model is illustrated in Figure 1. The 40-dimensional log-mel filterbank (Fbank40) were extracted every 10ms with 25ms of frame length. The output of MagicNet neural network is speech present probability for each frame. Generally, there is a post-processing to reduce noise in raw results from the neural network, but here we ignore it and focus on the architecture of neural network.

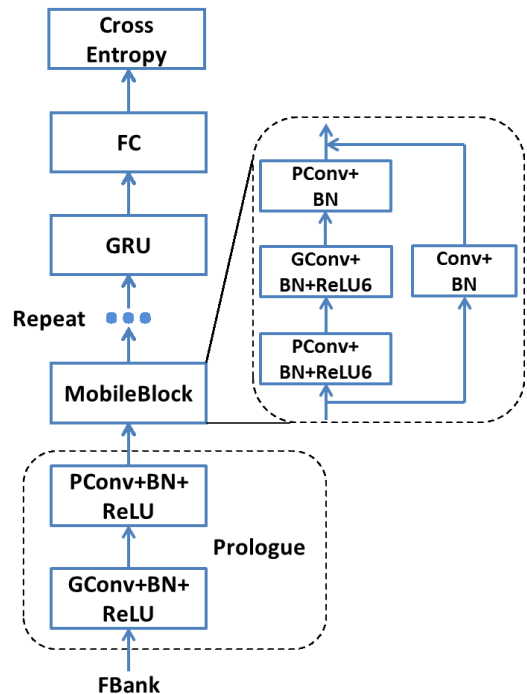

Figure 1: *The diagram of MagicNet model*

The CLDNN architecture has been used in many speech tasks for its more powerful ability in modeling feature variations than individual CNN, RNN and DNN. So we present a CLDNN-like neural network, named MagicNet, which is a cascade of MobileBlocks, a GRU, and a feed-forward neural network. A diagram of the MagicNet is shown in Figure 1. The convolution layers include a prologue block and two Mobile Blocks. The prologue block is a time-channel separable convolution, which consists of a group convolution (GConv) and a point convolution (PConv) each with a batch normal (BN) layer and a ReLU activation function. The Mobile Block is an inverted residual structure with linear bottleneck appeared in the MobileNet neural network but here 2 dimension convolutions are replaced by 1 dimension convolutions for less computation. All convolutions in the model are casual means that no future frames are used for real-time and low latency processing. The LSTM are replace by the GRU for same reasons. The last block is a 1 layer fully connected (FC) neural network for mapping hidden embedding to speech present probability. The detailed parameters of MagicNet are shown in Table 1.

Table 1: *The detailed parameters of MagicNet*

| Block | Layer | Output channel | Kernel | Stride |
|---|---|---|---|---|
| Prologue | GConv | 40 | 40 | 2 |
|  | PConv | 20 | 1 | 1 |
| Mobile Block 1 | Pconv | 20 | 1 | 1 |
|  | Gconv | 80 | 41 | 2 |
|  | Pconv | 20 | 1 | 1 |
| Mobile Block 2 | Pconv | 20 | 1 | 1 |
|  | Gconv | 80 | 21 | 2 |
|  | Pconv | 20 | 1 | 1 |
| GRU | Layers_num=2, Hidden_size=20 | | | |
| FC | Layers_num=1, Output_size=1 | | | |

## 3. Experiments

### 3.1. Training Dataset

In this work, the training dataset consists of a great part of the Chinese corpuses of AISHELL-3 [17] and all of QUT-NOISE [18] database. The corpuses of AISHELL-3 contains roughly 85 hours of audio spoken by 218 native Chinese mandarin speakers and total 88035 utterances recorded in quiet rooms. The QUT-NOISE database includes ten scenes of double channel noise about 14 hours. First, in consideration of imbalance between speech and non-speech in the dataset, utterances of each speaker were concatenated with a silent interval of random duration in 2 to 5 seconds. Then frame-level labels of speech and non-speech were obtained by Kaldi tools and CVTE Mandarin model v2. Labeled audio recording were split into 68 hours and 17 hours for training and testing respectively.

### 3.2. Training Method

Generally, noise-robust VAD systems are developed using audio from clean speech datasets augmented with different types of noise. The clean speech dataset were augmented by adding noise from QUT-NOISE database in the SNR of 5dB to 30dB and reverberation from sox tools in the probability of about one-third. Augmented audio were converted to segments of length 20s for simulating the classical scene of far-field voice interaction. We pre-processed those segments with Fbank40 features for training. One tenth of those samples were used for validation when training. All models were trained by the Adam optimizer with a 0.0001 initial learning rate. We trained all models on 8 Tesla K80 GPUs with a batch size of 50 per GPU until new minimum loss didn't appear during 50 epochs. All models were implemented and trained by the PyTorch toolkit.

### 3.3. Evaluation Method

To evaluate the performance of MagicNet VAD, the classical Google WebRTC vad[7] and two state-of-the-art vad algorithms proposed in CRNN [19] and MarbleNet [20] are used as our baselines. The rest of AISHELL-3 dataset and the all of THCHS30 [21] dataset were applied to construct in-domain and out-domain test dataset respectively. Three types of unseen noises, named babble, car and machine, from the non-speech dataset of UTDallas [22, 23] were selected to corrupt the clean speech by SNR of 0dB. For comparison, the metrics of F1-score and AUC were selected as performance measurements. F1-score take into account both accuracy and recall metrics, which is a common evaluation index of binary classification problems. AUC is the area under receiver operating characteristic (ROC) curves, which is an overall metric for binary classification. To evaluate the real time performance of four methods, we compared their real time factor on a Intel(R) Xeon(R) E5-2690 v4 @ 2.60GHz CPU.

### 3.4. Results and Discussion

The number of parameters and real time factor(RTF) of four models are shown in Table *2*. We observe that the number of parameters of proposed method is only a quarter of the MarbleNet method and a little more than half of the CRNN method. Among neutral network based methods, proposed method has a best real time performace.

The results of evaluation on in-domain test dataset are presented in Table *3*. We repeated each trial 5 times for each Model to avoid the perturbation of random factors. We observe that the proposed MagicNet network outperforms all baseline algorithms in terms of both F1-score and AUC. The fact demonstrate that as the MagicNet method has less amount of parameters than two baseline methods it do show a better fitting capacity than the others in the matched dataset.

The results of evaluation on out-domain test dataset are presented in Table *4*. The improved performance is found for car and machine noise in terms of both F1-score and AUC. On babble noise, the baselines are relatively robust, but the proposed get a good average performance. The results demonstrate that the MagicNet neural network are more robust to unmatched data than the baseline algorithms.

Table 2: *The number of parameters and RTF of four models*

| Model | WebRTC | MarbleNet | CRNN | MagicNet |
|---|---|---|---|---|
| Parameters/K | \ | 85.2 | 36.8 | 22.7 |
| RTF | 0.015 | 0.038 | 0.045 | 0.034 |

Table 3: *Comparison of the WebRTC(RTC), MableNet (MB), CRNN (CR) and MagicNet (MG) on in-domain test set. B, C, M, and AVE means Babble, Car, Machine and Average respectively.*

| Metric | Type | B | C | M | AVE |
|---|---|---|---|---|---|
| F1-score | RTC | 0.603 | 0.584 | 0.555 | 0.581 |
|  | MB | 0.823 | 0.850 | 0.818 | 0.830 |
|  | CR | 0.813 | 0.837 | 0.789 | 0.813 |
|  | MG | **0.833** | **0.882** | **0.862** | **0.859** |
| AUC | RTC | 0.719 | 0.701 | 0.653 | 0.691 |
|  | MB | 0.935 | 0.942 | 0.922 | 0.933 |
|  | CR | 0.934 | 0.948 | 0.909 | 0.930 |
|  | MG | **0.948** | **0.967** | **0.952** | **0.956** |

Table 4: *Comparison of the WebRTC(RTC), MableNet (MB), CRNN (CR) and MagicNet (MG) on **out-domain** test set. B, C, M, and AVE means Babble, Car, Machine and Average respectively.*

| Metric | Type | B | C | M | AVE |
|---|---|---|---|---|---|
| F1-score | RTC | 0.730 | 0.733 | 0.725 | 0.729 |
|  | MB | 0.872 | 0.887 | 0.852 | 0.870 |
|  | CR | **0.873** | 0.891 | 0.864 | 0.876 |
|  | MG | 0.866 | **0.906** | **0.872** | **0.881** |
| AUC | RTC | 0.674 | 0.679 | 0.618 | 0.657 |
|  | MB | 0.927 | 0.939 | 0.897 | 0.921 |
|  | CR | **0.930** | 0.950 | 0.910 | 0.930 |
|  | MG | 0.926 | **0.957** | **0.926** | **0.936** |

## 4. Conclusions

In this paper, we present the MargicNet, a computation-efficient and low-latency deep Voice Activity Detection model. We evaluate the MagicNet on an in-domain and an out-domain test dataset. We demonstrate that MagicNet is able to achieve state-of-the-art performance with the small RTF compared to the baseline neural network models. This ability of the proposed method is beneficial to the deployment of VAD on the computationally constrained environments such as mobile and wearable devices.

## 5. Acknowledges

We would like to thank the Haier U-home AI team for the help and valuable feedback